# Solving the Schrodinger Equation of an Electron in a Periodic Crystal Potential through Elliptic Functions

**Luca Nanni**

Faculty of Natural Science, University of Ferrara, 44122 Ferrara, Italy

luca.nanni@edu.unife.it

https://orcid.org/0000-0002-5121-5736

**Abstract**

*In this study, the Schrodinger equation of a valence electron in a periodic crystal potential is formulated and solved using the elliptic function formalism. The method allows double-periodic lattice planes to be represented in the Gauss plane. The reality of the obtained eigenfunctions and the structure of the valence and conduction bands are also investigated.*

## 1. Introduction

Elliptic functions are considered a special class of analytic mathematical functions that are used to investigate and solve problems—mainly in physics and astronomy—whose symmetry presents a double periodicity [1–5]. More specifically, the equations of motion of the system contain an expression dealing with the arc of an ellipse. Elliptic functions allow an exact solution to be found to problems that would otherwise present serious difficulties if approached with other analytical and algebraical tools [6–8]. Although the theory of elliptic functions has been used for more than a century in various scientific disciplines, it remains largely unknown and unused in chemistry. However, solid state chemistry is an ideal field for applying elliptic functions due to the periodicity of crystal lattices .

In this study, we will prove that the periodic potential generated by the ions of a metal crystal lattice, whose analytical formulation often leads to operators of complex form that make it difficult, if not impossible, to exactly solve the Schrodinger equation of the system, can be expressed as a Weierstrass elliptic function. This approach is possible if we consider the crystal lattice as an infinite repetition of a lattice plane along a given crystallographic direction. The lattice plane, itself formed by the infinite repetition of the unit cell (i.e., a two-dimensional Bravais lattice), represents the doubly periodic 2-

manifold. This makes the application of the theory of elliptic functions suited to describe the physicochemical quantities we want to investigate. The Schrodinger equation of a valence electron of a crystal is thus formulated using the periodic potential in the form of an elliptic operator, resulting in an equation which, regardless of the geometry of the unit cell, is always possible to solve exactly. In particular, the eigenfunctions of the Hamiltonian operator are Bloch waves whose periodic component is represented by a suitable elliptic function. Finally, by exploiting the fact that the electron wave vector appearing in the Bloch function must be real, it is possible to construct the valence and conduction bands of the crystal.

The article is organised as follows: in Sect. 2 we give some basic notions of Weierstrass elliptic functions and state some theorems necessary to set up the problem and solve the associated Schrodinger equation. In Sect. 3, the periodic potential is formulated as a Lamé function of order one, from which the Schrodinger equation of the problem is constructed. In Sect. 4 the Schrodinger equation is solved exactly, resulting in Bloch waves represented by quasi-periodic elliptic functions. The reality of the elliptic eigenfunctions is also discussed. Finally, in Sect. 5, by studying the analytical properties of the solutions, we obtain information on the valence and conduction bands that characterise the electronic structure of the metal.

## 2. Notions of elliptic functions

An elliptic function is defined as a doubly periodic meromorphic function [1]:

$$f(z + 2\omega_1) = f(z) \text{ and } f(z + 2\omega_2) = f(z) : \omega_1, \omega_2 \in \mathbb{C} \text{ and } \frac{\omega_1}{\omega_2} \ni \mathbb{R}. \quad (1)$$

Therefore, the elliptic functions are defined in a two-dimensional lattice of the Gauss plane, whose unit cell is given by the set $\{0, 2\omega_1, 2\omega_2, (2\omega_1 + 2\omega_2)\}$. The complex numbers $\omega_1$ and $\omega_2$ are called fundamental periods. Once the function within the cell is known, it remains defined for every other point of its definition domain $A \subset \mathbb{C}$.

*Definition* 1: two points $z_1, z_2 \in \mathbb{C}$ are congruent if they satisfy the following relation:

$$z_1 - z_2 = 2m\omega_1 + 2n\omega_2 \quad \forall m, n \in \mathbb{Z}.$$

From Definition 1, it is clear that $f(z_1) = f(z_2)$.

*Definition* 2: a set of points $z_i \in \{0, 2\omega_1, 2\omega_2, (2\omega_1 + 2\omega_2)\}$, where $f(z)$ is not defined, is called a set of irreducible poles.

The number of poles, each counted with its multiplicity, defines the order of the elliptic function.

*Theorem* 1: the sum of the residues of an elliptic function of an irreducible set of poles is zero:

$$\sum_{z_i} Res(f, z_i) = \frac{1}{2\pi i} \oint_{\partial Cell} f(z) dz,$$

where $\partial_{Cell}$ is the frontier (perimeter) of the cell $\{0, 2\omega_1, 2\omega_2, (2\omega_1 + 2\omega_2)\}$. From Theorem 1 it is clear that the order of an elliptic function must be equal to or greater than two. It follows that an elliptic function with an empty set of irreducible poles is a constant function.

The simplest elliptic function is that which has one pole per cell with multiplicity two. This function can be written as an infinite sum of functions $1/z^2$ each shifted by $(2m\omega_1 + 2n\omega_2)$:

$$\wp(z) = \frac{1}{z^2} + \sum_{m,n \neq \{0,0\}} \left[ \frac{1}{(z + 2m\omega_1 + 2n\omega_2)^2} - \frac{1}{(2m\omega_1 + 2n\omega_2)^2} \right]. \quad (2)$$

As usual $m, n \in \mathbb{Z}$. Equation (2) is called the Weierstrass elliptic function and is an even function with pole $z = 0$ of multiplicity two. It satisfies the Weierstrass differential equation [9]:

$$\left( \frac{d\wp(z)}{dz} \right)^2 = 4\wp^3(z) - g_2 \wp(z) - g_3,$$

where $g_2$ and $g_3$ are complex numbers that can be obtained from the series expansion of the Weierstrass function [10]. Being an even function, $\wp(z)$ is stationary at half-period.

Let us now introduce the function $\zeta(z)$, defined as:

$$\zeta(z) = \frac{1}{z} + \sum_{m,n \neq \{0,0\}} \left[ \frac{1}{z + 2m\omega_1 + 2n\omega_2} - \frac{1}{2m\omega_1 + 2n\omega_2} + \frac{z}{(2m\omega_1 + 2n\omega_2)^2} \right]. \quad (3)$$

Equation (3) is odd and has a single pole $z = 0$ with multiplicity one. Moreover, $\zeta'(z) = -\wp(z)$, $\zeta(z + 2\omega_1) = \zeta(z) + 2\zeta(\omega_1)$ and $\zeta(z + 2\omega_2) = \zeta(z) + 2\zeta(\omega_2)$. For these reasons Eq. (3) is not an elliptic function and is called a quasi-periodic function. This function also satisfies the following relation:

$$\omega_2 \zeta(\omega_1) - \omega_1 \zeta(\omega_2) = \frac{\pi i}{2}.$$

Finally, let us introduce the function $\sigma(z)$, defined as:

$$\sigma(z) = z \prod_{m,n \neq \{0,0\}} \frac{z + 2m\omega_1 + 2n\omega_2}{2m\omega_1 + 2n\omega_2} e^{\left( -\frac{z}{2m\omega_1 + 2n\omega_2} + \frac{z^2}{2(2m\omega_1 + 2n\omega_2)^2} \right)}. \quad (4)$$

Equation (4) is an odd analytic function whose periodicity is:

$$\sigma(z + 2\omega_i) = -\sigma(z) e^{2\zeta(\omega_i)(z + \omega_i)}, \quad i = 1,2$$

Therefore, $\sigma(z)$ is also a quasi-periodic function.

*Theorem 2*: let $f(z)$ be an elliptic function with a set $\{u_i\}$ of irreducible poles, each with multiplicity $r_i$. Then it can be expressed in terms of the quasi-periodic function $\zeta(z)$:

$$f(z) = K + \sum_i \sum_{j=i} \frac{(-1)^{j-1} c_{i,j}}{(j-1)!} \frac{d^{(j-1)} \zeta(z - u_i)}{dz^{(j-1)}}.$$

Here, $K$ is a numerical constant and $c_{i,j} = Res\left(f, u_i^{(j)}\right)$, where $u_i^{(j)}$ stands for the i[th] pole with multiplicity $j$.

*Theorem 3*: let $f(z)$ be an elliptic function with a set $\{u_i\}$ of irreducible poles, each with multiplicity $r_i$, and let $\{w_j\}$ be a set of irreducible roots of the equation $f(z) = 0$, each with multiplicity $s_j$. Then, $f(z)$ can be expressed in terms of the quasi-periodic function $\sigma(z)$:

$$f(z) = K \frac{\prod_j \sigma^{s_j}(z - w_j)}{\prod_i \sigma^{r_i}(z - u_i)},$$

where $K$ is a numerical constant.

To complete this section, we write down the addition formulas for the Weierstrass and quasi-periodic functions:

$$\begin{cases} \wp(z + \omega) = -\wp(z) - \wp(\omega) + \frac{1}{4}\left(\frac{\wp'(z) - \wp'(\omega)}{\wp(z) - \wp(\omega)}\right)^2 \\ \wp(z) - \wp(\omega) = -\frac{\sigma(z - \omega)\sigma(z + \omega)}{\sigma^2(z)\sigma^2(\omega)} \\ \zeta(z + \omega) - \zeta(z) - \zeta(\omega) = \frac{1}{2}\frac{\wp'(z) - \wp'(\omega)}{\wp(z) - \wp(\omega)} \end{cases} \quad (5)$$

All of the above notions will be used to develop the theory that is the subject of this article.

## 3. Periodic crystal potential

As is well known, the potential generated by atomic nuclei *seen* by a valence electron of a metal varies slowly within the unit cell [11, 12]. This is due to the fact that the innermost electrons shield the ions attenuating their true potential. This potential is called the pseudo-potential and represents a non-local operator, depending not only on the lattice coordinate but also on the atomic wave function. For a metal, this potential reads [13]:

$$U_{eff}(\mathbf{r}) = -\frac{Z}{|\mathbf{r}|} + \sum_\alpha (E_n - E_\alpha)\psi_\alpha(\mathbf{r}) \int \psi_\alpha^*(\mathbf{r}') e^{i\mathbf{k}\cdot\mathbf{r}'} d\mathbf{r}', \quad (6)$$

where $U_{eff}$ is the pseudo-potential, $Z$ is the atomic number, $E_n$ is the energy of the state corresponding to the quantum number $n$, $E_\alpha$ is the energy of the $\alpha$[th] bound atomic electron, $\psi_\alpha(\mathbf{r})$ is its wave function and $\mathbf{k}$ is the wave vector. The use of the potential in the form of Eq. (6) may entail mathematical complications depending on the complexity of the crystal lattice. Things change, however, if the potential is expressed through Lamé functions [14]. These functions are elliptic functions solving the Lamé equation:

$$\frac{d^2 y(\omega)}{dx^2} = \bigl(\alpha + \beta \wp(\omega)\bigr) y(\omega), \tag{7}$$

where $\alpha$ is numerical constant, $\beta = l(l+1)$ with $l$ a positive integer, and $\wp(\omega)$ is a Weierstrass elliptic function. Setting $l = 1$ to obtain a first order Lamé function, Eq. (7) can be rewritten as follows:

$$\left[-\frac{1}{2}\frac{d^2}{d\omega^2} + 2\wp(\omega)\right] y(\omega) = E y(\omega), \tag{8}$$

which is the dimensionless Schrodinger equation of an electron in a double-periodic potential $U = 2\wp(\omega)$. For clarity, we set $-\alpha/2 = E$; i.e., the energy of the eigenstates $y(\omega)$. This means that for the problem we are studying, the constant $\alpha$ must be real.

The idea of representing the periodic potential with Lamé functions is justified by the fact that these functions are suited to studying systems whose equations of motion differ from the classical ones for small fluctuations, and this, as already mentioned above, is precisely what characterises the pseudo-potential in a metal [15]. Solving Eq. (8) means finding the wave functions of the Bravais lattice that characterises a given crystallographic plane and the energies of the valence electrons. The form of these functions is expected to be analogous to that of Bloch waves [11, 12].

## 4. Solving the Schrodinger equation

As anticipated in the previous section, the solutions of Eq. (8) must be Bloch waves with double periodicity. The Bloch eigenfunction reads as follows:

$$\psi(\boldsymbol{r}) = u(\boldsymbol{r}) e^{i\boldsymbol{k}\cdot\boldsymbol{r}}, \tag{9}$$

where $u(\boldsymbol{r})$ must have the same periodicity as the crystal. Let us now introduce the following linearly independent elliptic functions:

$$y_\pm(\omega, a) = \frac{\sigma(\omega \pm a)}{\sigma(\omega)\sigma(\pm a)} e^{-\zeta(\pm\gamma)\omega}, \tag{10}$$

where $\zeta$ and $\sigma$ are the quasi-periodic functions defined in Eq. (3) and (4), respectively, $\omega$ is the complex variable, $\gamma$ is a constant that will be clarified later and, finally, $a$ is a parameter running on the frontier $\partial(0, \omega_1, \omega_2, \omega_3)$. This means we are studying the two-dimensional Bravais lattice in the Gauss plane, and the translational periodicity of the primitive cell is generated by $2\omega_1$ and $2\omega_2$ (which correspond to the unit cell vectors in the real plane). In the case of centred cells, the lattice node in the centre will have periodicity $\omega_3 = (\omega_1 + \omega_2)$.

Let us now substitute Eq. (10) into Eq. (8); the second derivative is:

$$\frac{d^2 y_\pm(\omega, a)}{d\omega^2} = \{[\zeta(\omega \pm a) - \zeta(\omega) \mp \zeta(a)]^2 - \wp(\omega \pm a) + \wp(\omega)\} y_\pm(\omega, a). \tag{11}$$

Using the first and third of the addition formulas in Eq. (5), the term $\wp(\omega \pm a)$ becomes:

$$\wp(\omega \pm a) = -\wp(\omega) - \wp(\pm a) + \frac{1}{4}\left(\frac{\wp'(\omega) - \wp'(\pm a)}{\wp(\omega) - \wp(\pm a)}\right)^2 \quad (12)$$
$$= -\wp(\omega) - \wp(\pm a) + \left\{\frac{1}{2}[\zeta(\omega \pm a) - \zeta(\omega) \pm \zeta(\pm a)]\right\}^2.$$

Substituting Eq. (12) into Eq. (11), Eq. (8) returns the eigenvalue $E = -\wp(\pm a) = -\wp(a)$, as the Weierstrass function is an even function. Therefore, the energy of the valence electron depends only on the complex variable $a$. Since, as already said, the pseudo-potential varies little in the unit cell, and since $U = 2\wp(\omega)$, it follows that the energy of the valence electron also has a value that differs little from one point to another in the cell. In this regard, if we substitute Eq. (10) into the Weierstrass equation and set the first term equal to zero, we obtain the following third order equation:

$$4\wp^3(\omega) - g_2\wp(\omega) - g_3 = 0, \quad (13)$$

which once solved gives the stationary points of the elliptic function. As mentioned previously, Eq. (13) is satisfied when $\wp(\omega)$ takes values at half period $\omega_1, \omega_2, \omega_3$. We denote by $e_1, e_2, e_3 = \wp(\omega_1), \wp(\omega_3), \wp(\omega_2)$ the three roots of such an equation. Since the eigenvalue of energy is given by $E = -\wp(a)$, it follows that its stationary values are given by $E_i = -e_i$ in the regime of $\omega = 0$ and that they must be real numbers. The three roots satisfy the following mutual relations [1]:

$$\begin{cases} e_1 + e_2 + e_3 = 0 \\ e_1e_2 + e_2e_3 + e_1e_3 = -\dfrac{g_2}{4} \\ e_1e_2e_3 = \dfrac{g_3}{4} \end{cases}. \quad (14)$$

The complex numbers $g_2$ and $g_3$ have already been introduced in Sect. 2, and their explicit form is:

$$\begin{cases} g_2 = 60 \sum\limits_{m,n \neq \{0,0\}} \left[\dfrac{1}{(2m\omega_1 + 2n\omega_2)^4}\right] \\ g_3 = 140 \sum\limits_{m,n \neq \{0,0\}} \left[\dfrac{1}{(2m\omega_1 + 2n\omega_2)^6}\right] \end{cases}. \quad (15)$$

Therefore, substituting Eq. (15) into Eq. (14) and solving the latter as a usual system of algebraic equations, we obtain the explicit forms of $e_i$ and therefore also of the energies corresponding to the stationary points of the elliptic solution of the Schrodinger equation. From the first of the relations in Eq. (14), it is clear that not all energies can have the same sign. Furthermore, once the chemical nature of the ions of the crystal is set, the energy is determined only by the geometry of the unit cell, and this is exactly what is expected if we recall the *classical* quantum theory of metals.

We have therefore proved that the elliptic functions in Eq. (10) are solutions of the Schrodinger equation of the system, with eigenvalues given by $-\wp(a)$. In a more compact form, the dimensionless Schrodinger equation can be written as $\mathcal{H}y_\pm(\omega, a) = -\wp(a)y_\pm(\omega, a)$, where $\mathcal{H} = d^2/2\omega^2 + 2\wp(\omega)$.

We now have all the information necessary to investigate the band structure of the metal. However, it is first necessary to study the reality of the obtained eigenfunctions: this property, in fact, will facilitate the study of the collective behaviour of the crystal's valence electrons.

## 5. On the reality of functions $y_\pm(\omega, a)$

Let us consider the rectangle of vertices $\{0, \omega_1, \omega_2, \omega_3\}$, which we will denote by $\partial\{0, \omega_1, \omega_2, \omega_3\}$, where $\omega_1$ lies on the real axis and $\omega_2$ lies on the imaginary one (we are always in the complex plane). Moreover, we suppose that $\wp(a)$ is always real (a necessary condition considering that energy is an intrinsically real quantity) as are the three roots $e_1, e_2, e_3$. As usual the parameter $a$ runs on the perimeter of this rectangle. Since the elliptic function $\wp(\omega)$ is even and of order two, there is always a point $b$ of the unit cell such that $\wp(a) = \wp(b)$. Under these hypotheses, the functions $y_\pm(\omega, a)$ can be shown to be real valued by verifying whether the equality $\bar{y}_\pm = y_\pm$ holds or not. The following cases hold:

$$\begin{cases} 1) \; a \in [0, \omega_1], a = b \text{ and } \wp(a) = \wp(b) > e_1 & \Rightarrow \; \bar{y}_\pm(\omega, b) = y_\pm(\omega, b) \\ 2) \; a \in [0, \omega_2], a = ib \text{ and } \wp(a) < e_3 & \Rightarrow \; \bar{y}_\pm(\omega, ib) = y_\mp(\omega, ib) \\ 3) \; a \in [\omega_2, \omega_3], a = \omega_2 + b \text{ and } e_3 < \wp(a) < e_2 & \Rightarrow \; \bar{y}_\pm(\omega, \omega_2 + b) = y_\pm(\omega, \omega_2 + b) \\ 4) \; a \in [\omega_1, \omega_3], a = \omega_2 + ib \text{ and } e_2 < \wp(a) < e_1 & \Rightarrow \; \bar{y}_\pm(\omega, \omega_2 + ib) = y_\mp(\omega, \omega_2 + ib) \end{cases}.$$

The first two cases can be trivially proved using the definitions in Eq. (3) and (4) of quasi-periodic functions. The cases 3) and 4) are proved by applying the conditions of quasi-periodicity of the functions $\sigma(\omega)$ and $\zeta(\omega)$ [1]:

$$\begin{cases} \zeta(\omega, \omega_i + b) = \zeta(\omega, \omega_i) + 2\zeta(b) \\ \sigma(\omega, \omega_i + b) = -\sigma(\omega, \omega_i) e^{\zeta(\omega, b)(\omega_i + b)} \end{cases},$$

where $i = 1, 2, 3$.

Let us now consider the case in which one root is real and the other two are complex. More precisely we suppose that $e_2$ is real while $e_1$ and $e_3$ are complex conjugates. Then, the following two cases hold:

$$\begin{cases} 1) \; a \in [0, \omega_1 + \omega_2], a = b \text{ and } \wp(a) > e_2 & \Rightarrow \; \bar{y}_\pm(\omega, b) = y_\pm(\omega, b) \\ 2) \; a \in [0, \omega_1 - \omega_2], a = ib \text{ and } \wp(a) < e_2 & \Rightarrow \; \bar{y}_\pm(\omega, ib) = y_\mp(\omega, ib) \end{cases}.$$

Both cases can be proved using the definitions in Eq. (3) and (4).

## 6. The electronic band structure

The quantum theory of metals predicts that the solutions of the Schrodinger equation for valence electrons are Bloch waves [11, 12]. Therefore, we must transform the functions in Eq. (10) to make them analogous to Bloch eigenfunctions. Supposing that the periodicity of the pseudo-potential is $2\omega_i$ and using the condition of quasi-periodicity of the function $\sigma(\omega)$, we can rewrite the first part of Eq. (10) as:

$$\frac{\sigma(\omega \pm a + 2\omega_i)}{\sigma(\omega + 2\omega_i)\sigma(\pm a)} = \frac{-e^{\zeta(\omega_i)(\omega \pm a + \omega_i)}\sigma(\omega \pm a)}{-e^{2\zeta(\omega_i)(\omega + \omega_i)}\sigma(\omega)\sigma(\pm a)} \\ = \frac{\sigma(\omega \pm a)}{\sigma(\omega)\sigma(\pm a)} e^{\pm 2a\zeta(\omega_i)}, \tag{16}$$

Supposing that all the roots $e_1, e_2, e_3$ are real, then Eq. (10) can be written as follows:

$$y_\pm(\omega, a) = f_\pm(\omega, a) e^{\pm i\lambda(a)\omega}, \tag{17}$$

where:

$$\begin{cases} f_\pm(\omega, a) = \dfrac{\sigma(\omega \pm a)}{\sigma(\omega)\sigma(\pm a)} e^{\mp \frac{a\zeta(\omega_i)}{\omega_i}\omega} \\ i\lambda(a) = \dfrac{a\zeta(\omega_i) - \omega_i\zeta(a)}{\omega_i} \end{cases}. \tag{18}$$

It is easy to verify that $y_\pm(\omega + 2\omega_i, a) = y_\pm(\omega, a)$, namely the eigenfunction in Eq. (17) has the typical periodicity of a Bloch wave. If we compare the second term of Eq. (18) with the argument of the exponential function (plane wave) of the Bloch wave (see Eq. (9)), we recognise that $\lambda(a) = -i(a\zeta(\omega_i) - \omega_i\zeta(a))/\omega_i$ corresponds to (but is not equal to) the dot product $\boldsymbol{k} \cdot \boldsymbol{r}$ in the real two-space. Therefore, to have physical meaning, the number $\lambda(a)$ must be purely imaginary. For this purpose, we note that the function $(a\zeta(\omega_i) - \omega_i\zeta(a))$ is a quasi-periodic function:

$$\begin{cases} (a\zeta(\omega_i) - \omega_i\zeta(a))\big|_{(a+2\omega_1)} = (a\zeta(\omega_i) - \omega_i\zeta(a))\big|_{(a)} \\ (a\zeta(\omega_i) - \omega_i\zeta(a))\big|_{(a+2\omega_2)} = (a\zeta(\omega_i) - \omega_i\zeta(a))\big|_{(a)} + i\pi \end{cases}. \tag{19}$$

Moreover, the following equalities hold:

$$\begin{cases} (a\zeta(\omega_i) - \omega_i\zeta(a))\big|_{(\omega_1)} = 0 \\ (a\zeta(\omega_i) - \omega_i\zeta(a))\big|_{(\omega_2)} = i\pi/2 \end{cases}. \tag{20}$$

Equations (19) and (20) can be easily proved using the definition of $\zeta(\omega_i)$. Calculating the derivative of the function $(a\zeta(\omega_i) - \omega_i\zeta(a))$, we get:

$$\frac{d}{da}(a\zeta(\omega_i) - \omega_i\zeta(a)) = \zeta(\omega_i) + \omega_i\wp(a). \tag{21}$$

This derivative is stationary along all the perimeter $\partial\{0, \omega_1, \omega_2, \omega_3\}$. Therefore, using Eq. (19) and (20), together with Eq. (21), we find that:

$$\begin{cases} a \in [0, \omega_1] \cup [\omega_2, \omega_3] \Rightarrow Re[\zeta(\omega_i) + \omega_i\wp(a)] \neq 0 \text{ and } Im[\zeta(\omega_i) + \omega_i\wp(a)] = 0 \\ a \in [0, \omega_2] \cup [\omega_1, \omega_3] \Rightarrow Re[\zeta(\omega_i) + \omega_i\wp(a)] = 0 \text{ and } Im[\zeta(\omega_i) + \omega_i\wp(a)] \neq 0 \end{cases}.$$

Based on this analysis we can conclude that:

$$\begin{cases} 1)\ a \in [0, \omega_1] \Rightarrow \text{only } Re(a\zeta(\omega_i) - \omega_i\zeta(a)) \neq 0 \text{ and monotonically increasing from } -\infty \text{ to } 0 \\ 2)\ a \in [\omega_1, \omega_2] \Rightarrow \text{only } Im(a\zeta(\omega_i) - \omega_i\zeta(a)) \neq 0 \text{ and monotonically increasing from } 0 \text{ to } i\pi/2 \\ 3)\ a \in [\omega_2, \omega_3] \Rightarrow Im(a\zeta(\omega_i) - \omega_i\zeta(a)) = i\pi/2 \text{ and } Re(a\zeta(\omega) - \omega\zeta(a)) = 0 \text{ at } \omega = \omega_2, \omega_3 \\ 4)\ a \in [0, \omega_2] \Rightarrow \text{only } Im(a\zeta(\omega_i) - \omega_i\zeta(a)) \neq 0 \text{ and monotonically increasing from } -i\infty \text{ to } i\pi/2 \end{cases}.$$

Therefore, $\lambda(a)$ is purely imaginary only within the set $[0, \omega_2] \cup [\omega_1, \omega_2]$ and hence the eigenfunction Eq. (17) is a Bloch wave. In conclusion, if $e_1, e_2, e_3$ are real, then the band structure of the valence electrons is formed by a finite number of levels comprised between the energies $-e_1$ and $-e_2$, while the conduction band is formed by an infinite number of levels with energy greater than $-e_3$.

Let us now consider the case in which $e_2$ is real and $e_1, e_3$ are imaginary. Then, the periodicity of the pseudo-potential is $(2\omega_1 + 2\omega_2)$ and $\lambda(a) = (a\zeta(\omega_1 + \omega_2) - (\omega_1 + \omega_2)\zeta(a))$. Repeating the analysis done for the previous case, we obtain:

$$\begin{cases} (a\zeta(\omega_1 + \omega_2) - (\omega_1 + \omega_2)\zeta(a))\big|_{(a+2\omega_1)} = (a\zeta(\omega_1 + \omega_2) - (\omega_1 + \omega_2)\zeta(a))\big|_{(a)} - i\pi \\ (a\zeta(\omega_1 + \omega_2) - (\omega_1 + \omega_2)\zeta(a))\big|_{(a+2\omega_2)} = (a\zeta(\omega_1 + \omega_2) - (\omega_1 + \omega_2)\zeta(a))\big|_{(a)} + i\pi \\ (a\zeta(\omega_1 + \omega_2) - (\omega_1 + \omega_2)\zeta(a))\big|_{(\omega_1+\omega_2)} = 0 \\ (a\zeta(\omega_1 + \omega_2) - (\omega_1 + \omega_2)\zeta(a))\big|_{(\omega_1-\omega_2)} = -i\pi \end{cases}. \quad (22)$$

Moreover:

$$\frac{d}{da}\left(a\zeta(\omega_1 + \omega_2) - (\omega_1 + \omega_2)\zeta(a)\right) = \zeta(\omega_1 + \omega_2) + (\omega_1 + \omega_2)\wp(a). \quad (23)$$

Using Eq. (22) and (23), we conclude that $\lambda(a)$ is purely imaginary only within the set $[0, \omega_1 - \omega_2]$ and the electron structure is formed only by a conductive band made of an infinite number of levels with energies greater than $-e_2$.

## 7. Concluding discussion

In this article, the Schrodinger equation of a valence electron in a periodic potential has been formulated and solved using the formalism of elliptic functions. This application is made possible by the fact that the lattice planes of the crystal are characterised by a double periodicity and can be represented in the Gauss plane. This is the domain in which elliptic functions find their natural application. In this framework the pseudo-potential, which must have the same periodicity as the crystal lattice, is expressed with a Weierstrass elliptic function, thus obtaining an operator which, once inserted in the Schrodinger equation, allows the problem to be solved by obviating the complexity that is encountered when a *classical* analytical approach is used [16, 17]. The solutions are elliptic eigenfunctions which, with appropriate manipulations, can be transformed in such a way as to have an analytical form completely analogous to the usual Bloch waves. From the analytical comparison between these two functions and considering that the electron wave vector must be a real quantity, it is possible to determine the sets in which the elliptic eigenfunction is defined within the unit cell and, consequently, to calculate the valence and conduction bands and their energies.

This study suggests that some problems of quantum chemistry characterised by multiple periodicity, which typically have laborious and often only approximate solutions, can be revisited in the formalism of elliptic functions in order to simplify as much as possible the complexity of the mathematical approach. For example, some problems that might benefit from this approach are the study of the chemical-physical properties of metals and semiconductors, or the Fourier analysis for the resolution of the structure of crystals by X-ray and neutron diffraction, for which only some timid and unfinished work appears in the literature [18–20]. We believe that these gaps are attributable only to the lack of knowledge of the theory of elliptic functions in the chemical-physical field and of their considerable potential in solving complex problems.

**Declarations**

The author certify do not have affiliations with or involvement in any organization or entity with any financial interest (such as honoraria; educational grants; participation in speakers' bureaus; membership, employment, consultancies, stock ownership, or other equity interest; and expert testimony or patent-licensing arrangements), or non-financial interest (such as personal or professional relationships, affiliations, knowledge or beliefs) in the subject matter or materials discussed in this manuscript.